\documentclass[showpacs,showkeywords,twocolumn,superscriptaddress,amssymb]{revtex4-1}
\usepackage{bm,color,amsmath,amssymb,mathrsfs,latexsym,graphicx,psfrag,hyperref,bbm,float}

\newcommand{\ket}[1]{|#1\rangle}
\newcommand{\ev}[1]{\langle #1 \rangle}

\newcommand{\be}[0]{\mathbf e}

\newcommand{\bq}[0]{\mathbf q}
\newcommand{\bd}[0]{\mathbf d}
\newcommand{\bk}[0]{\mathbf k}
\newcommand{\bR}[0]{\mathbf R}
\newcommand{\bG}[0]{\mathbf G}

\newcommand{\bp}[0]{\mathbf p}

\newcommand{\bl}[0]{{\boldsymbol \ell}}
\newcommand{\bu}[0]{\mathbf u}
\newcommand{\bv}[0]{\mathbf v}
\newcommand{\bx}[0]{\mathbf x}
\newcommand{\by}[0]{\mathbf y}

\newcommand{\bsigma}[0]{\boldsymbol \sigma}

\newcommand{\im}[0]{\mbox{Im}}
\newcommand{\mH}[0]{\mathcal{H}}
\newcommand{\mA}[0]{\mathcal{A}}
\newcommand{\mP}[0]{\mathscr{P}}
\newcommand{\mQ}[0]{\mathcal{Q}}

\newcommand{\bra}[1]{\langle #1|}

\newcommand{\spl}[1]{\begin{align}\begin{split} #1 \end{split} \end{align}}
\newcommand{\al}[1]{\begin{align} #1 \end{align}}
\newcommand{\Hbh}{H_{\mbox{\tiny BH}}}
\newcommand{\Hmf}{H_{\mbox{\tiny mf}}}
\newcommand{\mD}[0]{\mathcal{D}}

\newcommand{\Ncut}[0]{N_{\mbox{\tiny cut}}}
\newcommand{\Hqp}{H_{\mbox{\scriptsize QP}}}
\newcommand{\cre}[2]{#1^{\dagger}_{#2}}
\newcommand{\ann}[2]{#1^{\phantom{\dagger}}_{#2}}
\newcommand{\vvec}[2]{\begin{pmatrix} #1 \\ #2 \end{pmatrix}}

\newcommand{\tHqp}{\tilde{H}_{\mbox{\scriptsize QP}}}

\newcommand{\pic}[3]{
\begin{figure}[ht!]
\includegraphics[width=#3]{#1} \caption  {#2}
\end{figure}
}

\newcommand{\picwide}[3]{
\begin{figure*}[t!]
\includegraphics[width=#3]{#1} \caption{#2}
\end{figure*}
}


\begin{document}

\title{Quasi-Particle Theory for the Higgs Amplitude Mode} 
\author{Ulf Bissbort}
\affiliation{MIT-Harvard Center for Ultracold Atoms, Research Laboratory of Electronics, and Department of Physics,
Massachusetts Institute of Technology, Cambridge, Massachusetts 02139, USA}
\affiliation{Singapore University of Technology and Design, 138682 Singapore}
\affiliation{Institut f\"ur Theoretische Physik, Goethe-Universit\"at, 60438 Frankfurt/Main, Germany}
\author{Michael Buchhold}
\affiliation{Institut f\"ur Theoretische Physik, Goethe-Universit\"at, 60438 Frankfurt/Main, Germany}
\affiliation{Institut f\"ur Theoretische Physik, Universit\"at Innsbruck, A-6020 Innsbruck, Austria}
\author{Walter Hofstetter}
\affiliation{Institut f\"ur Theoretische Physik, Goethe-Universit\"at, 60438 Frankfurt/Main, Germany}

\begin{abstract}

We present a generalized quasi-particle theory for bosonic lattice systems, which naturally contains all relevant collective modes, including the Higgs amplitude in the strongly correlated superfluid. In contrast to Bogoliubov theory, this non-perturbative method does not rely on a small condensate depletion and is valid for any interaction strength in three spatial dimensions. It is based on an expansion around the bosonic Gutzwiller ground state in terms of appropriately chosen fluctuation operators and lays the foundation for the description of real-time dynamics in terms of the natural, weakly interacting quasi-particles. Furthermore, it provides a systematic framework for efficiently calculating observables beyond the Gutzwiller approximation and for including external perturbations, as well as higher order decay and interactions in terms of quasi-particle operators. It allows for the construction of an alternative path integral approach in terms of quasi-particle coherent states.
\end{abstract}
\maketitle

\picwide{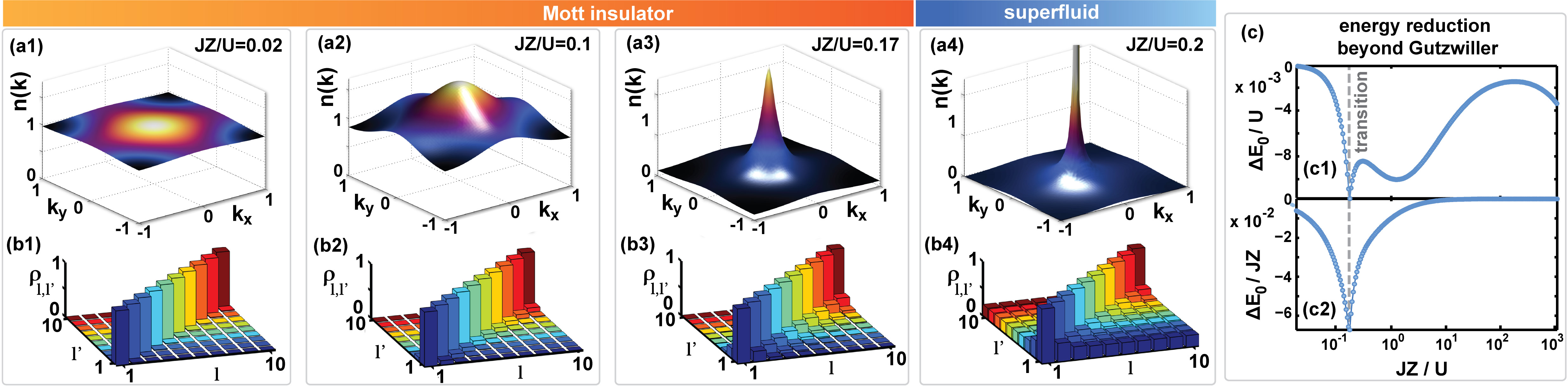}{\label{Fig1}Static observables calculated beyond GW mean-field: Quasi-momentum distributions $n(\bk)$ are shown in (a1)-(a4) for the 3D Bose-Hubbard model ($n=1$ and in the $k_z=0$ plane) for various values of $JZ/U$ with the corresponding single-particle density matrices $\rho_{\bl, \bl'}=\ev{b_\bl^\dag b_{\bl'} }$ in the real-space representation. In contrast to GW theory, $n(\bk)$ features a non-trivial structure in the MI originating from the presence of correlated particle-hole fluctuations in the ground state. $\rho_{\bl,\bl'}$ only depends on $\bl-\bl'$ and decays to zero exponentially in the MI with an increasing coherence length as the transition is approached. In the SF it approaches a constant value indicating off-diagonal long-range order. In (c) the ground state energy reduction $\Delta E_0$ relative to the GW energy is shown at constant filling $n=1$ as a function of $JZ/U$ (note the logarithmically scale). This is maximal at the SF-MI transition and vanishes for both $JZ\to 0$ and $U\to 0$ when scaled in the appropriate units.}{\linewidth}

After its recent experimental observation in strongly correlated superfluids (SF) based on lattice modulation \cite{Schori04,Endres12} \footnote{Note that the amplitude mode's signal in \cite{Schori04} was not identified in the work.} and Bragg spectroscopy\cite{bissbort11}, the Higgs amplitude mode (HAM) has been in the focus of a renewed wave of investigation. The HAM is an additional well-defined quasi-particle (QP) appearing in the SF of a 3D lattice \cite{altman02,Huber2007,Huber2008,menotti08,Krutitsky2011}, but was also recently shown to be robust against decay in 2D \cite{zwerger04,podolsky11,Endres12,pollet12,podolsky12,chen13,ruegg08,matsunaga13}. Generally, the HAM can appear when a continuous symmetry is spontaneously broken and an additional, independent quasi-particle mode emerges \cite{Anderson1963,Higgs1964}, for which the order parameter performs mainly amplitude oscillations under a coherent excitation. For strongly interacting bosons in a lattice, this can be understood to occur within a $U(1)$-symmetry breaking picture at the SF - Mott insulator (MI) transition. The theoretical description of the HAM thus inherently requires a theory valid at strong interactions, which is a more challenging task on the microscopic level than the perturbative inclusion of interaction effects, as in Bogoliubov theory (BT)\cite{Bogoliubov1947}. The latter has been highly successful in describing weakly interacting condensates both with and without a lattice. While our theoretical approach is applicable to general interacting bosonic lattice models, here we focus on the simplest case, given by the Bose-Hubbard Hamiltonian
\spl{
\label{EQ:BH_Hamiltonian}
\Hbh&=-J \sum_{\ev{\bl,\bl'}} b_\bl^\dag b_{\bl'}^{\phantom{\dag}}  - \mu \sum_\bl b_\bl^\dag b_{\bl}^{\phantom{\dag}} + \frac U 2 \sum_\bl b_\bl^\dag b_\bl^\dag b_{\bl}^{\phantom{\dag}} b_{\bl}^{\phantom{\dag}}.
}
This can be experimentally realized with remarkable control and probing capabilities in systems of ultracold bosonic atoms in the lowest band of an optical lattice \cite{bloch08}, where $J,\mu,U, \bl$ denote the nearest-neighbor tunneling, chemical potential, interaction energy and site index respectively. Here, we present a microscopic derivation of an extended quasi-particle theory on an operator level, valid throughout the entire phase diagram, which is non-perturbative in $U$ and contains all elementary excitations on equal footing, allowing for a calculation of observables beyond BT or Gutzwiller (GW) theory (shown in Fig.~\ref{Fig1}).
Apart from BT losing its validity, since the appearance of the HAM is related to a significant condensate depletion, an analysis of the number of Bogoliubov modes rules out the hope that any renormalization of BT may describe the HAM. The crucial constituent in constructing our QP theory at arbitrary interactions is a judicious choice of the fluctuation operators in terms of which the expansion around a suitable mean-field state is performed. Analogously to BT being the natural extension of Gross-Pitaevskii (GP) theory by systematically diagonalizing the Hamiltonian expanded to second order in fluctuation operators $\delta b_\bl =b_\bl-\ev{b_\bl}$ around the classical GP state (valid at $J\ll U$), here we perform an expansion around the variational bosonic GW ground state $\ket{\psi_{\mbox{\tiny GW}}}=\prod_{\otimes \bl} \ket{\psi_\bl}_\bl$ in a suitable set of quasi-bosonic fluctuation operators $\{\sigma , \, \sigma^\dag \}$, generalizing upon the MI case considered in \cite{altman03, navez13}. Just as BT can be derived by quantizing the eigenmodes of the linearized time-dependent GP equation \cite{fetter72} (up to the ambiguity of ordering higher order terms prior to quantization \cite{zinn2010}), our theory can be seen as the quantized counterpart of the theory obtained by linearizing the time-dependent GW equations of motion \cite{Krutitsky2011,bissbort11,natu2011}, which is known to naturally contain the HAM. In a path integral formulation, this can be regarded as a saddle point expansion around the bosonic GW state, as done for fermions in \cite{kotliar86}.\\
The GW ground state is determined by minimizing the variational energy $\bra{\psi_{\mbox{\tiny GW}}} \Hbh \ket{\psi_{\mbox{\tiny GW}}}$. We begin by defining the local fluctuation \textit{annihilation} operators $\tilde \sigma_\bl^{(i)}=\ket{0}_\bl {_{\bl}}\bra{i}\otimes \prod_{\bl'\neq \bl} \mathbbm 1_{\bl'}$ which, in the product basis of Wannier Fock states, map the local $i$-th  excited GW state $\ket i_\bl$ at site $\bl$ on the local GW ground state $\ket 0_\bl$. These operators $\tilde \sigma_\bl^{(i)}$ do not directly obey bosonic commutation relations, a problem also encountered in the spatial representation of particle and hole excitations in the MI and commonly remedied by imposing hardcore constraints \cite{altman02,altman03,Huber2007,Huber2008}. We choose a different approach and, with the position of lattice site $\bl$ being $\bR_\bl$, define the associated fluctuation operators in quasi-momentum space $\sigma_\bk^{(i)}=\frac{1}{\sqrt L} \sum_\bl e^{-i \bk \cdot \bR_\bl} \, {\tilde \sigma_\bl^{(i)}}$. These are approximately bosonic \vspace{-3mm}
\spl{
\label{EQ:GW_fluc_op_commutator}
[ { \sigma_{\bk'}^{(j)}}, \,  \sigma_\bk^{{(i)}^\dag} ]&=\delta_{\bk,\bk'} \, \delta_{i,j}-\frac 1 L R_{\bk,\bk'}^{(i,j)},
}
for low fluctuation \emph{densities}, as $\ev{R}$ (see \cite{supplement}(A)) scales linearly with the number of excitations. $\ev R /L$ is thus the small parameter of the theory, analogous to the condensate depletion in GP theory, and the consistency of the theory can be verified a posteriori. Since the GW eigenstates $\ket{i}_\bl$ constitute a local basis, the local operators $ {\tilde \sigma_{\bl}^{{(j)}^\dag}} \tilde \sigma_{\bl}^{{(i)}}=\ket j_\bl {_\bl} \bra i   , \; \tilde \sigma_{\bl}^{{(i)}}=\ket 0_\bl {_\bl} \bra i, \; {\tilde \sigma_{\bl}^{{(i)^\dag}}} = \ket i_\bl {_\bl} \bra 0$, and $[\mathbbm 1- \sum_{i>0} {\tilde \sigma_{\bl}^{{(i)^\dag}}} \tilde \sigma_{\bl}^{{(i)}}]=\ket 0_\bl {_\bl} \bra 0$ for $i>0$ form a complete local operator basis. Consequently, any global many-body operator can be expressed as a superposition of products of the local fluctuation operators $\tilde \sigma_{\bl}^{{(i)}}$, a property which carries over to the momentum fluctuation operators ${ \sigma_{\bk'}^{(j)}}$ by virtue of the Fourier transformation's unitarity. Specifically, $\Hbh$ can be expressed exactly in terms of $\sigma$-operators $\Hbh=E^{(0)}+\mH^{(2)}+\mH^{(3)}+\mH^{(4)}$
and first order terms vanish identically iff the operator expansion is performed around the classical GW ground state. $\mH^{(3)}$ and $\mH^{(4)}$ describe QP decay and interaction processes (see \cite{supplement}(C)) and are at most of order four, since $\Hbh$ couples at most two sites. Note that in contrast to BT, $\mH^{(3)}$ and $\mH^{(4)}$ are proportional to $J$. By definition $\mH^{(2)}$ is quadratic in the fluctuation operators and, using Eq.~(\ref{EQ:GW_fluc_op_commutator}) for $\ev R \ll L$ can be written as a quadratic form
\spl{
\label{EQ:H2_GW_fluc_matrix_form}
\mH^{(2)}=& \frac 1 2
\begin{pmatrix}  
{\boldsymbol\sigma}\\ {\boldsymbol\sigma}^\dag
\end{pmatrix}^\dag \, \Hqp 
		\begin{pmatrix}  
{\boldsymbol\sigma}\\ {\boldsymbol\sigma}^\dag
\end{pmatrix} - \frac 1 2 \mbox{Tr} \, h 
}
with the coefficient matrix $\Hqp$, the matrix elements of which can be directly obtained \cite{supplement}(B). For states containing a low density of QP, the higher order terms are less relevant and the dynamics is primarily governed by $\mH^{(2)}$. The QP basis $\mH^{(2)}$ is diagonal in terms of newly defined bosonic operators $\beta$ and $\beta^\dag$. 
In order to preserve bosonic commutation relations, the $H_{QP}$ has to be diagonalized according to a generalized, Bogoliubov type diagonalization, see \cite{supplement}(D).
At this point we have to distinguish between two qualitatively different cases: in the Mott insulator $\Sigma \mH^{(2)}$ is diagonalizable \footnote{$\Sigma=\mbox{diag}(\mathbbm{1},-\mathbbm{1})$ defines the metric on the symplectic space, inherent to bosonic systems.} and a complete basis of eigenvectors exists; in the condensate this is however no longer true and one can at most bring the $\Sigma \mH^{(2)}$ into Jordan normal form (see \cite{supplement}(D)). Physically, this is related to the property that no bosonic QP mode can be attributed to the orbital single-particle condensate mode. Generally, the eigenvectors of $\Sigma \mH^{(2)}$ appear in pairs $\bx^{(s)}=(\bu^{(s)}, -\bv^{(s)})^t$ and $\by^{(s)}=(-\bv^{(s)^*}, \bu^{(s)^*})^t$ for conjugate non-zero eigenvalues $\omega_s$ and $-\omega_s^*$ respectively. Imposing the normalization $\bx^{(s)^\dag} \Sigma \bx^{(s)}=1$ and $\by^{(s)^\dag} \Sigma \by^{(s)}=-1$, the corresponding QP operators are explicitly given in terms of the original fluctuation operators by \vspace{-3mm}
\spl{
\label{EQ:QP_operators}
\beta_s={\bx^{(s)}}^\dag \Sigma \vvec{{\bsigma}}{{\bsigma}^\dag} \hspace{2mm} \mbox{ and } \hspace{2mm}  \beta_s^\dag=-{\by^{(s)}}^\dag \Sigma \vvec{{\bsigma}}{{\bsigma}^\dag},
}

\vspace{-3mm} \noindent
which also approximately fulfill $[\beta_s,\beta_{s'}^\dag]=\delta_{s,s'}$. In the MI, these constitute a complete operator basis and an arbitrary many-body operator can be expressed in terms of $\beta_s$ and $\beta_{s'}^\dag$ or powers thereof. In the condensate, there is also one eigenvector $\bp$ to an eigenvalue zero $\Sigma \Hqp \, \bp=0$ and we furthermore implicitly define the generalized eigenvector $\bq$ and mass $\tilde m$ by $\Sigma \Hqp \, \bq = -\frac{i}{\tilde m} \bp$ and, since for both $\bp^\dag \Sigma \bp=\bq^\dag \Sigma \bq=0$, choose the euclidean normalization condition $\bq^\dag \bq=\bp^\dag \bp=1$, as well as the phase relation $\bq^\dag  \Sigma \bp=i$ \cite{blaizot1986}. The operators $\mathscr P={\bp}^\dag \Sigma \vvec{\tilde{\bsigma}}{\tilde{\bsigma}^\dag}$ and $\mathcal Q ={\bq}^\dag \Sigma \vvec{\tilde{\bsigma}}{\tilde{\bsigma}^\dag}$ associated with these are not of a bosonic nature, but resemble the position and momentum of an effective free particle of mass $\tilde m$ \cite{blaizot1986,lewenstein96} (see Fig.~\ref{fig2} (a) and (b)). Since $\Sigma \Hqp$ is not diagonalizable in the condensate and the eigenvectors are not orthogonal with respect to the euclidean norm, the direct representation in terms of eigenvectors cannot be used. Rather, the unit operator is expressed in terms of the vectors $\{ \bx^{(s)}, \by^{(s)}, \bp, \bq\}$ (see \cite{supplement}(D)) and the completeness relation is inserted to obtain
\spl{
\label{EQ:QP_Hamiltonian_condensate}
	\mH^{(2)}
= \sum_{s} \omega_{s}  \beta_{s}^\dag \, \beta_{s}  +\frac{\mathscr{P}^2}{2 \tilde m}
+ \Delta E_0.
}
in the condensate. The collective index $s$ denotes the mode (such as the HAM trap eigenmodes in \cite{Endres12}), which is reducible $s=(\bk,\gamma)$ into a quasi-momentum $\bk$ and internal mode index $\gamma$ for homogeneous systems. In the SF $\gamma=1$ is the Bogoliubov sound mode with a linear dispersion relation $\omega_{\bk,\gamma}=c|{\bf k}|$ for small $|\bk|$, where $c$ defines the speed of sound, which is strongly suppressed and deviates from the Bogoliubov result $c=\sqrt{2n_0UJ}$ for intermediate and strong interactions, as shown in Fig.~\ref{fig2}(d). This is of significance for entropy and particle transport in the SF shells of the typical wedding cake structure of a trapped BEC. On the other hand, as the SF-MI transition is approached, the effect of the HAM QPs becomes increasingly important for the calculation of observables and gains increasing amount of spectral weight, as seen in the composition of the single-particle density of states (DOS) $\rho(\omega)=-\sum_\bk \im \bra{\psi_0} a_\bk [\omega-\mH + i 0^+]^{-1} a_\bk^\dag \ket{\psi_0}/\pi$ shown in Fig.~\ref{fig:DOS}. Throughout the SF phase, the HAM has a non-zero energy gap $\Delta=\lim_{|{\bf k}|\rightarrow0}\omega_{{\bf k},\gamma=2}$, which vanishes exactly at the transition to the MI. \\
In the MI, on the other hand, the two lowest excitation modes $\gamma=1,2$ are the particle and hole modes with their order depending on $\mu$. Here, the MI QP-Hamiltonian is identical to Eq.~(\ref{EQ:QP_Hamiltonian_condensate}) up to the ${\mathscr{P}^2}/{2 \tilde m}$ term being absent. In both the SF and MI further higher bosonic modes also exist (e.g. multi-particle excitations in the MI). Whereas the dynamics of the SF is captured by an effective relativistic (non-relativistic) action for the order parameter for strong interactions close to the MI transition (weakly interacting SF) and the crossover between these two qualitatively different regimes is an interesting open question for future study, the higher order modes are never contained in such an effective theory for the order parameter alone. We remark that in our approach, the equations of motion govern the dynamics of the full local quantum state, from which the order parameter can be determined at any time. Our approach also sets the stage for the study of non-equilibrium dynamics in 3D, including the QP decay and interactions effects, as can for example be induced by quenches \cite{kollath07}.

\pic{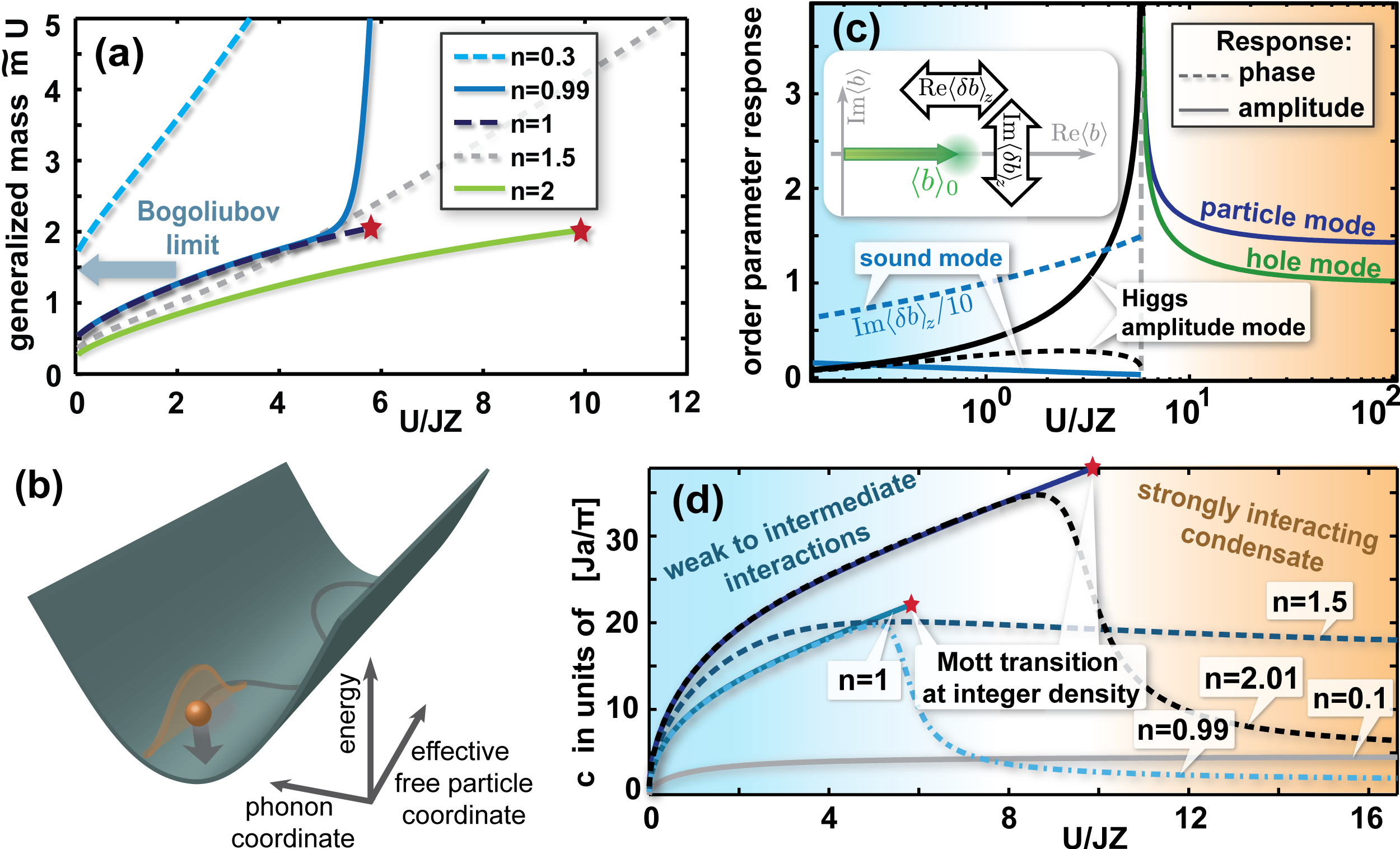}{\label{fig2}(Color online) In the classical limit, the dynamics of a lattice BEC can be pictured as the trajectory of an effective particle of mass $\tilde m$ in a high dimensional potential landscape, locally free along one coordinate (i.e. $\mQ$, for which the phase can drift and diffuse) and quadratic for each QP mode, as illustrated in (b). The mass $\tilde m$ of this fictitious particle and the speed of sound $c$ are shown in (a) and (d) respectively, and are strongly renormalized for $U\gg JZ$ compared to the BG result. For $U/JZ \to 0$ at constant density $U \ev{b}^2\rightarrow 0$ and the Bogoliubov result $\tilde{m}=\frac{1}{2U \ev b^2}$ is recovered. At integer density the system undergoes a transition into the MI, beyond which the sound and the effective free particle cease to exist. Note that $\tilde m U$ and $c/(Ja)$ remain finite at the transition ($a$ is the lattice spacing). In (c) the real and imaginary part of the order parameter response $\mbox{max}_\bl(\bra{z}\delta b_\bl \ket{z})/|z|$ is shown, where $\ket z=e^{-|z|^2/2 + z \beta_{\bk,\alpha}^\dag} \ket{\psi_0}$ is a coherent QP state at $\bk=  0.01\pi \be_x /a$ for $n=1$ and $\mu$ chosen such that $\omega_{\bk=0,\gamma=1}=\omega_{\bk=0,\gamma=2}$ in the MI. Note that the sound mode has a strong phase response and $\mbox{Im}\ev{\delta b}$ is scaled by a factor $0.1$ in (c). In the MI the response of the particle and the hole modes is $U(1)$-symmetric, hence $\mbox{Re}\ev{\delta b}=\mbox{Im}\ev{\delta b}$.}{\columnwidth}

We now focus on the eigenstates of $\mH^{(2)}$. The presence of the scalar energy reduction term (relative to the GW energy) $\Delta E_0 = -\frac 1 2 \left(\mbox{Tr}(h) - \sum_{s}  \omega_{s} \right)$ in Eq.~(\ref{EQ:QP_Hamiltonian_condensate}) and shown in Fig.~\ref{Fig1}(c) directly implies that the ground state of our theory differs from and improves the GW ground state $\ket{\psi_{\mbox{\tiny GW}}}$ if $\Delta E_0<0$. Indeed, the energy is reduced intermediate coupling shown in Fig.~\ref{Fig1}(c) and furthermore, as expected from the GW ansatz becoming exact in both the non-interacting and the strongly interacting limit, $\lim_{J\to 0} \Delta E_0 / U =\lim_{U\to 0} \Delta E_0 / J =0$ in the appropriate units. From the form of Eq.~(\ref{EQ:QP_Hamiltonian_condensate}) and $\beta_s$ being bosonic, it follows that the ground state of $\mH^{(2)}$ can be implicitly defined by
\vspace{-1mm}
\spl{
\label{EQ:gs_implicit}
\beta_{s} \ket{\psi_0}=0 \qquad \mathscr{P} \ket{\psi_0}=0
}

\vspace{-2mm} \noindent
for all modes $s$, i.e. it contains no QPs in the new QP basis. The excited QP Fock states can subsequently be constructed by adding QPs $\ket{\{n_s\}} \propto \prod_s (\beta_s^\dag )^{n_s} \ket{\psi_0}$. By decomposing a given initial state in the basis of these eigenstates of $\mH^{(2)}$, dynamical calculations can be performed. In the condensate, where $[\mathcal Q,\mathscr{P}]=i$ resemble position and momentum of a free particle in one dimension with periodic boundary conditions (i.e. an $O(2)$ rigid rotor) and the ground state in this sector fulfills $\mathscr{P} \ket{\psi_0}=0$. It is possible to express $\ket{\psi_0}$ explicitly in terms of the operators $\sigma$. However, to calculate the time-dependence and expectation values of various operators, it is convenient to first express all operators in terms of $\{\beta_s,\beta_s^\dag, \mathscr{P}, \mathcal Q \}$ and then use the commutation relations in conjunction with Eq.~($\ref{EQ:gs_implicit}$).

\pic{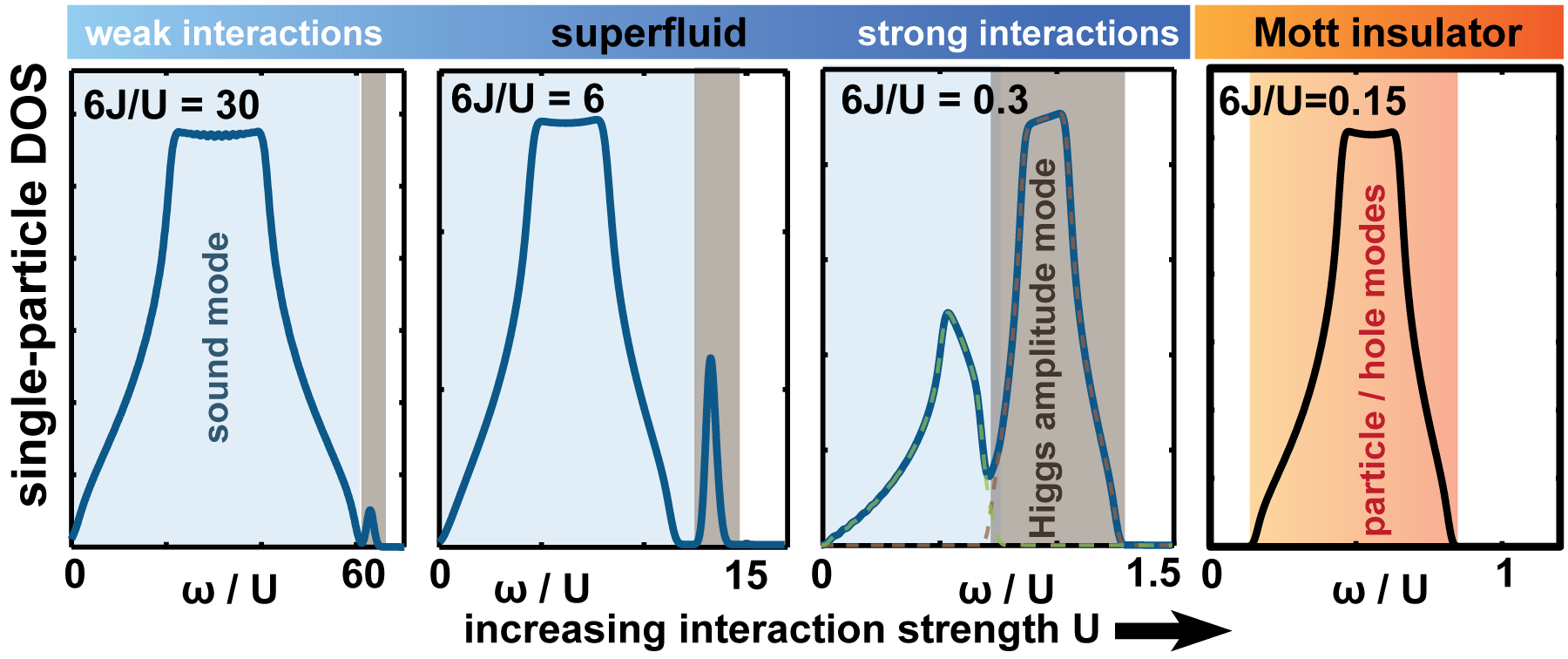}{\label{fig:DOS}(Color online) Positive frequency part of the single particle DOS ($n=1$) for different interaction strengths $U$. Note the evolution of the amplitude
mode with decreasing interaction strength: In the strongly correlated SF at $JZ/U=0.3$, it corresponds to the higher energy peak carrying more spectral weight than the sound
mode. With decreasing interaction strength, it continuously loses spectral weight
and transforms into a small peak just above the sound mode. The DOS in the MI (right panel) corresponds to the particle mode, possessing a finite gap in contrast to the SF.}{\linewidth}

The BT form $\mP=(e^{-i\phi} \delta a_0+e^{i\phi} \delta a_0^\dag)/\sqrt 2$ reveals that $\mP$ is associated with fluctuations in the amplitude (i.e. condensate density) of the SF order parameter with a phase $\phi$. Thus the second term in Eq.~\eqref{EQ:QP_Hamiltonian_condensate} energetically penalizes these amplitude fluctuations, which are decoupled from the phase fluctuations of the condensate mode described by $\mQ$ in the limit ${\bf k}\rightarrow 0$ \cite{blaizot1986, lewenstein96, villain97}. The smaller the effective mass $\tilde{m}$, the higher the fluctuation energy cost and it is interesting to note that although $\tilde{m}$ generally increases with $U$, it stays finite at the MI transition at integer density, as shown in Fig.~\ref{fig2}(a). The conjugate variable ($\mathcal{Q} \propto e^{-i\phi} \delta a_0-e^{i\phi} \delta a_0^\dag$ in BT) represents phase fluctuations of the SF order parameter, which cost no energy in the linearized regime, thus not appearing in Eq.~(\ref{EQ:QP_Hamiltonian_condensate}). However, the commutation relation $\dot{\mQ}=i{[H,\mQ]=\mathcal{P}/\tilde{m}}$ leads to non-trivial dynamics of $\mathcal{Q}$, describing the decay of long-range correlations in time
\spl{\label{EqCor}
C_{0}(t)=\langle \{a^{\phantom{\dagger}}_{\bk=0}(t),a^{\dagger}_{\bk=0}\}\rangle\propto \langle e^{-i\mathcal{Q}(t)}\rangle=e^{-\frac{t^2}{2\tilde{m}^2}\langle\mathcal{P}^2\rangle}.}
This decay describes an exponential collapse of coherence \cite{villain97}, which can be detected in cold atom experiments \cite{greiner02,will10} and is also naturally described for large $U/J$ beyond BT by our approach. To demonstrate the physical signatures of the most relevant QP modes, we calculate the order parameter phase and amplitude response $\delta\psi_l/|z|$ in a coherent QP state, shown in Fig.~\ref{fig2}(c). In the limit $|\bk|\to 0$ the response of the sound mode always becomes purely phase-like, however the response of the amplitude mode is generally not purely in the amplitude, but also contains a phase. Only in the limit of approaching the MI transition and $|\bk|\to 0$ its signature is a pure amplitude excitation.

In order to incorporate the effect of external perturbations and determine operator expectation values for the many-body system, it is crucial to express a given operator in the QP representation, i.e. in the operator basis $\mathcal B=\{ \cre{\beta}{{\bf k},\alpha}, \ann{\beta}{\bk, \alpha},\mP,\mQ\}$. This is performed by expressing the operator in terms of the fluctuation operators $\{\cre{\sigma}{},\sigma\}$ and subsequently in terms of $\mathcal B$ by using the inverse transformation of Eq.~\eqref{EQ:QP_operators}, derived in \cite{supplement}(E). This transformation is exact and the general form of an operator $\mathcal{A}_{\bk}$ in $\bk$-space up to linear order in $\mathcal B$ is \vspace{-1mm}
\spl{
\label{Eq:Op_trafo}
\mathcal{A}_{\bk}=\sum_{\gamma}\left[ D_{\bk,\gamma}\cre{\beta}{\bk,\gamma}+\tilde{D}_{-\bk,\gamma}\ann{\beta}{-\bk ,\gamma}\right]\hspace{-0.5mm}+\hspace{-0.5mm}D_P\mathcal P\hspace{-0.5mm}+\hspace{-0.5mm}D_Q \mathcal Q,
}

\vspace{-3mm} \noindent
where the coupling coefficients $D_{X}$ depend on the GW states and the transformation matrices (see \cite{supplement}(E)). Expressing an external operator acting on the system through Eq.~(\ref{Eq:Op_trafo}), the $D_s, \tilde D_s$ thus give an intuitive understanding of the various mode's significance for the respective process. Furthermore using Eq.~(\ref{Eq:Op_trafo}) in conjunction with Eq.~(\ref{EQ:gs_implicit}) directly allows for the calculation of static and dynamic expectation values in the QP ground state, circumventing the explicit calculation of the latter. 

Specifically, we calculate and show the momentum distribution $n(\bk)$ and the corresponding real-space single-particle density matrix $\rho_{\bl,\bl'}=\ev{b_\bl^\dag b_{\bl'}}$ across the SF-MI transition in Fig.~\ref{Fig1}(a) and (b). Compared to GW and Bogoliubov theory, $n(\bk)$ is strongly modified, revealing a non-trivial momentum distribution in the MI (where $\rho_{\bl,\bl'}$ decays exponentially with $|\bl-\bl'|$), which diverges for $\bk\rightarrow0$ at the transition.

This approach is also suitable to formulate an alternative coherent state path integral approach, which explicitly incorporates the sound and amplitude mode and additional higher order terms describing QP interactions. Defining local $\sigma$-coherent states $\ket{\Phi} = e^{\sum_{j>0} \phi_j \sigma^{(j)^\dag}} \ket{0}$, where $\Phi=(\phi_1,\phi_2,\ldots)$, the identity reads
\spl{
\mathbbm{1}=\int d\Phi \, d\Phi^* \, e^{-\sum_{j>0} |\phi_j|^2 } \ket{\Phi}\bra{\Phi}
}
with the measure $d\Phi \, d\Phi^*=\prod_{j>0} \frac{d\mbox{\scriptsize Re}\, \phi_j \; d\mbox{\scriptsize Im}\, \phi_j}{\pi}$. Although the $\sigma$-coherent states are not exact eigenstates, but rather $\sigma^{(j)} \ket{\Phi}=\phi_j \ket 0$, the eigenstate relation applies under the integral
\spl{
\sigma^{(j)} \mathbbm{1}= \int d\Phi \, d\Phi^* \, e^{-\sum_{j>0} |\phi_j|^2 } \phi_j \ket{\Phi}\bra{\Phi}.
}
This provides a starting point to construct an alternative path integral representation in the $\sigma$-coherent state basis. For instance, extending to $L$ sites, the action corresponding to $\Hbh$ is given by
\spl{
\mathcal S=\int_0^\beta d\tau \left[ \sum_{\bl,j}    \phi_j^{(\bl)} \, \partial_\tau \phi_j^{(\bl)} + H(\Phi, \Phi^*) \right],
} 
where in $H(\Phi, \Phi^*)$ is the energy function obtained from expressing $\Hbh=H(\boldsymbol \sigma, \boldsymbol \sigma^\dag)$ in normal order and $\beta$ is the inverse temperature.
This formalism thus gives rise to a non-relativistic field theory of first order in the time derivative, which contains the HAM. Calculating expectation values and correlations in this $\sigma$-path integral is equivalent to the QP theory in the operator formulation above. However, in contrast to the usual many-body coherent state path integral, the classical path corresponds to the GW state and provides a better starting point for a perturbative expansion and diagrammatics and an intuitive picture at strong correlations. In contrast to the path integral in the bare bosonic fields, this formulation contains the self-energy up to second order in the fluctuations. 

In conclusion, we have derived a bosonic QP theory for arbitrary interaction strengths, which describes all relevant excitations on an equal footing and recovers a number of existing approaches in the respective limits. Our theory provides an intuitive understanding of the dynamics of strongly interacting Bose systems in terms of optimal weakly interacting QPs, analogous to, but beyond BT.

\begin{acknowledgments}
We thank J.~Anglin, A.~Daley, A.~Fetter, V.~Liu and P.~Navez for helpful discussions.
This work was supported by the DFG via Forschergruppe FOR 801 and SFB TR 49.
\end{acknowledgments}

\bibliography{text}

\clearpage

\begin{appendix}

\section{Control Parameter}
The small parameter for the validity of our quasi-particle theory is not the condensate depletion (as would be the case for Bogoliubov theory), but a more abstract quantity. For the $\sigma$-operators to be bosonic, the expectation values of the operators
\spl{
\label{EQ:fluc_op_comm_rest}
R_{\bk,\bk'}^{(i,j)}&=\sum_\bl e^{i(\bk-\bk') \cdot \bR_\bl} \left(     
\tilde \sigma_{\bl}^{{(i)}^\dag}\,\tilde \sigma_{\bl}^{{(j)}} +\delta_{i,j} \sum_{j'>0} \tilde \sigma_{\bl}^{{(j')}^\dag}\, \tilde \sigma_{\bl}^{{(j')}}
 \right)\\
&= \sum_{\bk_1} \left(   \sigma_{\lfloor \bk_1+\bk-\bk' \rfloor}^{{(i)}^\dag}\, { \sigma_{\bk_1}^{(j)}} + \delta_{i,j}\sum_{j'>0}  \sigma_{\lfloor \bk_1+\bk-\bk' \rfloor}^{{(j')}^\dag} \, { \sigma_{\bk_1}^{(j')}} \right)
}
have to be small. Thus, a single parameter quantifying the validity of the theory is given by the expectation value of the operator $R/L=\sum_{i,j,\bk,\bk'} R_{\bk,\bk'}^{(i,j)}/L$ evaluated within the state of interest. Here $\lfloor \bk \rfloor$ denotes the addition of a suitable reciprocal lattice vector to the argument $\bk$, such that the enclosed quasi-momentum lies within the first Brillouin zone. This control parameter $\ev R$ is a measure the density of `fluctuations' beyond the GW state and is required to be small if the fluctuation expansion to be meaningful. 

For the three-dimensional lattice it is found to have a kink at the MI - SF phase transition where it takes on a global maximum value of $0.04$ and decays to zero in both the strongly interacting and non-interacting limits, where the GW description becomes exact. This justifies the theory in three dimensions and gives an estimation for the relative quantitative error for various calculable quantities at a given point in the phase diagram.

Note a fundamental difference to Bogoliubov theory: whereas the validity of Bogoliubov theory requires the depletion of the condensate to be small, the transformed Bogoliubov quasi-particle operators are exactly bosonic (assuming the original creation and annihilation operators are bosonic) at arbitrary interactions strengths exceeding the theory's realm of validity. This is not true for our quasi-particle theory, where the quasi-particle operators lose their bosonic character if the density of excitations becomes too large.

\section{Matrix elements of the quasi-particle Hamiltonian matrix}

\paragraph{Second order quasi-particle Hamiltonian}
All terms from $\Hmf$, as well as four different types of terms from $H_\delta$ (the four different combinations of creation and annihilation operators on two neighboring sites) contribute to the second order Hamiltonian $\mH^{(2)}$. The latter terms make it non-diagonal unless $J=0$. Defining the elements 
\al{
B_{i,j}^{(\bl)} &= {_\bl}\bra{i} b_\bl \ket{j}_\bl\\
\label{EQ:B_tilde_zero} \tilde B_{i,j}^{(\bl)} &=B_{i,j}^{(\bl)} - B_{0,0}^{(\bl)} \, \delta_{i,j}\\
F_{i_1,i_2,j_1,j_2}^{(\bl_1,\bl_2)}&= \tilde B_{j_1,i_1}^{(\bl_1)^*} \tilde B_{i_2,j_2}^{(\bl_2)} + 
 \tilde B_{i_1,j_1}^{(\bl_1)} \tilde B_{j_2,i_2}^{(\bl_2)^*},
}
which, by virtue of $\tilde B_{0,0}^{(\bl)}=0$, possess the properties
\al{
\label{EQ:F_coeff_zero_property}
	F_{0,i_2,0,j_2}^{(\bl_1,\bl_2)}&=F_{i_1,0,j_1,0}^{(\bl_1,\bl_2)}=0\\	 	
	\label{EQ:F_coeff_symm_property}
	F_{i_2,i_1,j_2,j_1}^{(\bl_2,\bl_1)}&={F_{j_1,j_2,i_1,i_2}^{(\bl_1,\bl_2)^*}}=F_{i_1,i_2,j_1,j_2}^{(\bl_1,\bl_2)}.
}
and using the property that one can rewrite 
\spl{
\ket{i}_{\bl_1} {_{\bl_1}}\bra{0} \otimes  \ket{0}_{\bl_2} {_{\bl_2}}\bra{j}=\tilde \sigma_{\bl_1}^{{(i)}^\dag} \tilde \sigma_{\bl_2}^{{(j)}}
}
and analogues thereof, we find
\spl{
\label{EQ:QP_H2_fluc_Wannier}
\mH^{(2)}
&=\sum_{i>0} \sum_\bl E_i^{(\bl)} \,  \tilde \sigma_\bl^{{(i)}^\dag} \,  \tilde \sigma_\bl^{(i)}\\& - J \sum_{i,j>0}  \sum_{\ev{\bl_1,\bl_2}}\left[ 2F_{i,0,0,j}^{(\bl_1,\bl_2)}  \, \tilde \sigma_{\bl_1}^{{(i)}^\dag} \tilde \sigma_{\bl_2}^{{(j)}} \right. \\
& + \left. F_{0,0,i,j}^{(\bl_1,\bl_2)} \, \tilde \sigma_{\bl_1}^{{(i)}} \tilde \sigma_{\bl_2}^{{(j)}} +F_{i,j,0,0}^{(\bl_1,\bl_2)} \, \tilde \sigma_{\bl_1}^{{(i)}^\dag} \tilde \sigma_{\bl_2}^{{(j)}^\dag} \right].
}
In grouping the terms we used $F_{i,0,0,j}^{(\bl_1,\bl_2)} = F_{0,i,j,0}^{(\bl_2,\bl_1)}$ from Eq.~(\ref{EQ:F_coeff_symm_property}). We define  $J_{\bl,\bl'}$ to be $J$ if sites $\bl$ and $\bl'$ are nearest neighbors and zero otherwise. Thus for the same site, $J_{\bl,\bl}$ always vanishes and using the property that $ \tilde \sigma_{\bl_1}^{{(i)}^\dag}$ and $\tilde \sigma_{\bl_2}^{{(j)}}$ always commute for neighboring lattice sites, we rewrite Eq.~(\ref{EQ:QP_H2_fluc_Wannier}) as
\spl{
\label{EQ:QP_H2_fluc_Wannier_symmetrized_fluc}
\mH^{(2)}
=&\sum_{i>0} \sum_\bl E_i^{(\bl)} \,  \tilde \sigma_\bl^{{(i)}^\dag} \,  \tilde \sigma_\bl^{(i)}- \frac 1 2 \sum_{i,j>0}  \sum_{\bl_1,\bl_2} J_{\bl_1,\bl_2}\\
\times & \left[ F_{i,0,0,j}^{(\bl_1,\bl_2)}  \, \tilde \sigma_{\bl_1}^{{(i)}^\dag}\tilde  \sigma_{\bl_2}^{{(j)}} + F_{i,0,0,j}^{(\bl_1,\bl_2)}\, \tilde \sigma_{\bl_2}^{{(j)}}  \, \tilde \sigma_{\bl_1}^{{(i)}^\dag}  \right. \\ & + \left. F_{0,0,i,j}^{(\bl_1,\bl_2)} \, \tilde \sigma_{\bl_1}^{{(i)}} \tilde \sigma_{\bl_2}^{{(j)}} +F_{i,j,0,0}^{(\bl_1,\bl_2)} \, \tilde \sigma_{\bl_1}^{{(i)}^\dag} \tilde \sigma_{\bl_2}^{{(j)}^\dag} \right],
}
which is a more suitable to express $\mH^{(2)}$ in matrix form. Note that a factor $\frac 1 2$ enters in the second term, since the sum $\sum_{\bl_1,\bl_2}$ contains twice as many non-vanishing terms as before. 

It is now possible to express the second order Hamiltonian in matrix form
\spl{
\label{EQ:H2_GW_fluc_matrix_form}
\mH^{(2)}=& \frac 1 2
\begin{pmatrix}  
\tilde{\boldsymbol\sigma}\\ \tilde{\boldsymbol\sigma}^\dag
\end{pmatrix}^\dag \, \Hqp 
		\begin{pmatrix}  
\tilde{\boldsymbol\sigma}\\ \tilde{\boldsymbol\sigma}^\dag
\end{pmatrix},
}
where
\begin{equation}
	\Hqp=	\begin{pmatrix}  
h& \Delta\\
\Delta^*& h^*
\end{pmatrix},
\end{equation}
is the coupling matrix in the Wannier representation and $\tilde{\boldsymbol\sigma}$ and $\tilde{\boldsymbol\sigma}^\dag$ are both column vectors of $\sigma_{\bl}^{(i)}$ and $\sigma_{\bl}^{{(i)}^\dag}$ operators respectively. In the general (potentially inhomogeneous) case the coefficient matrices $h$ and $\Delta$ are Hermitian and defined by their matrix elements
\spl{
h_{(i,\bl),(j,\bl')}&=\delta_{\bl,\bl'} \, \delta_{i,j} \, E_i^{(\bl)}-J_{\bl,\bl'} \, F_{i,0,0,j}^{(\bl,\bl')}\\
\Delta_{(i,\bl),(j,\bl')}&=- J_{\bl,\bl'}F_{i,j,0,0}^{(\bl,\bl')}.
}

\paragraph{Homogeneous System}
In a homogeneous system with a spatially constant potential and tunneling elements $J$, the structure of the coefficient matrices becomes particularly clear. 
For homogeneous systems, $\Sigma \mH^{(2)}$ is reducible to subblocks of dimension $2 L (\Ncut-1)$, where particle-like fluctuation operators of the $+\bk$ sector couple to hole-like fluctuations of the $-\bk$ sector. Given the non-interacting tight-binding dispersion relation
\begin{equation}
	\epsilon(\bk)\equiv 2 J \big[Z/2-\sum_{d=1}^{Z/2} \cos(\mathbf{a}_d \cdot \bk) \big],
\end{equation}
where $Z$ is the coordination number and $a_\bd$ is a real-space lattice vector, the coefficient matrix elements in the quasi-momentum representation are
\al{
\label{EQ:def_h_tilde_GW_QP}
\tilde h_{(i,\bk),(j,\bk')}&=\delta_{\bk,\bk'} \left[  \delta_{i,j} \, E_i +  (\epsilon(\bk) -JZ )  F_{i,0,0,j}\right]\\
\tilde \Delta_{(i,\bk),(j,\bk')}&= \delta_{-\bk,\bk'} \, (\epsilon(\bk) -JZ )\, F_{i,j,0,0}.
}
$E_i$ is the on-site energy of the $i$-th GW state, which is now site-independent for a homogeneous system, together with $F_{i,j,0,0}$ and $F_{i,0,0,j}$.
For the explicit numerical diagonalization, one has to truncate the local subspace to the $\Ncut$ lowest GW eigenstates, such that $\mH^{(2)}$ is $2 L (\Ncut-1)$-dimensional.

\section{Higher order terms}
The terms beyond second order in the fluctuation operators describe decay and interaction processes between the quasi-particles. Since the Bose-Hubbard Hamiltonian contains coupling terms between two sites at most, there are only terms of third and fourth order terms. In the basis of GW eigenstates, the third order terms are 
\spl{
\mH^{(3)}=&- J \sum_{ \stackrel{\ev{\bl_1,\bl_2}}{{i_1,i_2,i_3}>0}} \Big[ F_{0,i_1,i_2,i_3}^{(\bl_1,\bl_2)} \ket{0}_{\bl_1} {_{\bl_1}}\bra{i_2} \otimes  \ket{i_1}_{\bl_2} {_{\bl_2}}\bra{i_3}\\
&+F_{i_1,0,i_2,i_3}^{(\bl_1,\bl_2)} \ket{i_1}_{\bl_1} {_{\bl_1}}\bra{i_2} \otimes  \ket{0}_{\bl_2} {_{\bl_2}}\bra{i_3}\\
&+F_{i_1,i_2,0,i_3}^{(\bl_1,\bl_2)} \ket{i_1}_{\bl_1} {_{\bl_1}}\bra{0} \otimes  \ket{i_2}_{\bl_2} {_{\bl_2}}\bra{i_3} \\
&+F_{i_1,i_2,i_3,0}^{(\bl_1,\bl_2)} \ket{i_1}_{\bl_1} {_{\bl_1}}\bra{i_3} \otimes  \ket{i_2}_{\bl_2} {_{\bl_2}}\bra{0} \Big].
}
and for the fourth order term there is only one combination
\spl{
\mH^{(4)}=-J \hspace{-2mm}\sum_{ \stackrel{\ev{\bl_1,\bl_2}}{{i_1,i_2,j_1,j_2}>0}} F_{i_1,i_2,j_1,j_2}^{(\bl_1,\bl_2)} \ket{i_1}_{\bl_1} {_{\bl_1}}\bra{j_1} \otimes  \ket{i_2}_{\bl_2} {_{\bl_2}}\bra{j_2}.
}
Using the inverse of Eq.~\ref{EQ:QP_operators}, these can easily be expressed in terms of operators $\{\beta_s, \beta_s^\dag, \mP, \mQ \}$ and the amplitudes of the interaction and decay processes can be read off.

\section{Completeness Relations}
In contrast to the diagonalization of normal matrices ($[A,A^\dag]=0$), where a basis of eigenvectors always exists, it is not guaranteed that $\Sigma \Hqp$ is diagonalizable and always possesses a basis of eigenvectors. Although $\Hqp$ is Hermitian and thus normal, $\Sigma \Hqp$ is not. The non-diagonalizability is related to the existence of gapless Goldstone modes and the most general form a matrix can be transformed into is the Jordan normal form. Each Goldstone mode is related to the appearance of a Jordan subblock with generalized eigenvalue zero. In a physical picture, these degrees of freedom correspond to the dynamics of an effective free particle, whereas the dynamics in the eigenspace is that of harmonic oscillators. For the single species Bose-Hubbard model, we thus have to distinguish between two qualitatively different scenarios: in the MI, where no Goldstone mode exists, the matrix is diagonalizable. In the SF, however, the eigenvectors do not form a complete basis. Here we have to complete the basis with a generalized eigenvector, together with which the completeness relation (i.e. the identity matrix) can be formulated.

\subsection{Diagonalizable Case: Insulator}

If no eigenvalue vanishes, we have a complete basis of eigenvectors $\bx^{(\bk,\gamma)}$ and $\by^{(\bk,\gamma)}$. This structure also applies to each subblock $\bk\neq 0$ of $\Sigma \tHqp$ in the condensate. 
As can be checked using the orthogonality relations and the application to the basis set of eigenvectors, the unit operator can be written as
\spl{
\label{EQ:QP_completeness_non_condensate}
\mathbbm{1}^{(\bk)}&= \sum_{\gamma} (\bx^{(\bk,\gamma)} {\bx^{(\bk,\gamma)}}^\dag - \by^{(\bk,\gamma)} {\by^{(\bk,\gamma)}}^\dag)\Sigma.
}

\subsection{Non-Diagonalizable Case: Superfluid}
In the condensate regime, the sub-block corresponding to $\bk=0$ does not possess a basis of eigenvectors and this case, which leads to a qualitatively different Hamiltonian, has to be treated separately. This is not the case in the MI regime, where all eigenvalues of $\tHqp$ are non-zero and can be treated as described in the previous paragraph.

Let us denote the dimension of  $\tHqp$ by $2\mD$, where the limit $\mD\to \infty$ can be taken at the very end, when calculating any expectation values within this theory. If one zero eigenvalue of $\tHqp$ appears, one corresponding eigenvector of the structure 
\spl{
\label{EQ:GW_QP_p_eigenvec}
\bp=\vvec{\bu}{-\bu^*}
}
is guaranteed to exist, the $\Sigma$-norm of which vanishes exactly $\bp^\dag \Sigma \bp=0$. We choose to normalize it to unity with respect to the euclidean norm $\bp^\dag \bp=1$, which uniquely determines $\bp$ up to a overall complex phase. 

Assuming that only one pair of conjugated eigenvalues has become zero, the eigenvectors $\bx^{(\bk,\gamma)}$ and $\by^{(\bk,\gamma)}$ to all non-zero eigenvalues (i.e. $\gamma> 1$ by convention) span the $(2\mD-2)$-dimensional subspace and are mutually orthogonal and normalized as 
\al{
\bx^{(\bk,\gamma)^\dag} \Sigma \,\bx^{(\bk',\gamma')}&=\delta_{\bk,\bk'} \, \delta_{\gamma,\gamma'} \\
\by^{(\bk,\gamma)^\dag} \Sigma \,\by^{(\bk',\gamma')}&=- \delta_{\bk,\bk'} \, \delta_{\gamma,\gamma'} \\
\bx^{(\bk,\gamma)^\dag} \Sigma \,\by^{(\bk',\gamma')}&=0.
}
Since they are eigenvectors to different eigenvalues, $\bp$ is also orthogonal to and linearly independent of all other eigenvectors 
\al{
\bp^\dag \Sigma \bx^{(\bk,\gamma)}&=0\\
\bp^\dag \Sigma\, \by^{(\bk,\gamma)}&=0.
}
Hence, together with $\bp$, all eigenvectors span the entire space up to a single basis vector. Considering the eigenvector structure parametrically as a function of $\bk$ in the respective subblock, as $\bk$ approaches zero, the missing basis vector $\bq$ must lie in the two-dimensional subspace of the lowest eigenvalue pair for any non-zero $\bk$. One might therefore hope to construct it $\Sigma$-orthogonal to $\bp$ in this subspace. However, this is impossible since for $\bk=0$, it implies that the basis vector is a multiple of $\bp$. We could choose $\bq$ to lie anywhere within this limiting subspace as long as it is linearly independent of $\bp$. However, this would not allow for a direct representation of the identity in terms of eigenvectors and the vector $\bq$. The vector $\bq$ lies within the generalized subspace of the eigenvalue zero and can be chosen, such that its euclidean norm vanishes $\bq^\dag \Sigma \, \bq=0$. 
The application of $\Sigma \tHqp$ maps $\bq$ onto a multiple of $\bp$
\spl{
\label{EQ:mapping_q_onto_p}
\Sigma \tHqp \, \bq = -\frac{i}{\tilde m} \bp,
}
and we fix its normalization by requiring
\al{
\\
\bq^\dag  \Sigma \bp&=i.
}
The mass-like constant $\tilde m$ is real, which can easily be determined once $\bp$ and $\bq$ are known. The vector $\bq$ is of the form
\spl{
\label{EQ:GW_QP_q_form_fluc_chap}
\bq=-i\vvec{\bv}{\bv^*},
}
from which it is also directly clear that it has vanishing norm. Hence we see that in the basis spanned by all ($\bx^{(\bk,\gamma)},\; \by^{(\bk,\gamma)}, \; \bp,\; \bq$), the matrix $\Sigma \tHqp$ is in its Jordan normal form. We can now write the completeness relation in the condensate as
\spl{
\label{EQ:QP_completeness_condensate}
\mathbbm{1}=& \sum_{\bk,\gamma}' ( \bx^{(\bk,\gamma)} {\bx^{(\bk,\gamma)}}^\dag - \by^{(\bk,\gamma)} {\by^{(\bk,\gamma)}}^\dag)\Sigma \\
& + \; i (\bq \, \bp^\dag - \bp \, \bq^\dag)\Sigma,
}
where $\sum'$ denotes the summation over all combinations $(\bk,\gamma)$ except for the term $(\bk=0,\gamma=1)$.

\section{Inverse Transformation}
To express operators in terms of the QP operators, the inverse transformation is required, i.e. expressing $\{\sigma_{\bk,i}, \, \sigma_{\bk,i}^\dag \}$ in terms of $\{\beta_{\bk,\gamma}, \, \beta_{\bk,\gamma}^\dag \}$. The transformation in the MI and SF regimes are again qualitatively different and we discuss the more complicated, SF case. For the homogeneous system, the symmetry always allows us to fix the complex phases between related eigenvectors $\bx^{(\bk,\gamma)}=\bx^{(-\bk,\gamma)}$, $\by^{(\bk,\gamma)}=\by^{(-\bk,\gamma)}$  in the $\bk$ and $-\bk$ sectors, as well as all elements of the generalized eigenvectors to be real. Let us define the matrix $W$ containing the generalized eigenvectors as column vectors
\spl{
\label{EQ:GW_W_matrix_def_SF}
W=\Big[  \bx^{(1)}, \ldots ,\bx^{(\mD-1)},\,  i  \bp,\, \by^{(1)}, \ldots ,\by^{(\mD-1)},  i \bq  \Big],
}
for which one finds the relation
\spl{
\label{EQ:Sigma_tilde_def_GW}
W^\dag \Sigma W = 
\begin{pmatrix} 
\mathbbm{1}_{\mD-1} \\
 & 0  & \ldots &  -i\\
 &\vdots & -\mathbbm{1}_{\mD-1} &  \vdots \\
 & i  & \ldots &  0
\end{pmatrix} 
=: \tilde \Sigma.
}
The matrix $\tilde \Sigma$ is Hermitian and unitary and fulfills
\spl{
{\tilde \Sigma}^{-1}=\tilde \Sigma=\tilde \Sigma^\dag.
}  
Multiplying both sides of (\ref{EQ:Sigma_tilde_def_GW}) by ${\tilde \Sigma}$, one finds ${\tilde \Sigma} W^\dag \Sigma W = \mathbbm{1}_{2\mD}$, which allows the inverse matrix $W^{-1}$ to be obtained to be obtained by a simple matrix multiplication
\spl{
\label{EQ:GW_QP_def_W_inv}
W^{-1}={\tilde \Sigma} W^\dag \Sigma=
\begin{pmatrix} 
U^\dag, & V^\dag \\
-i {\bv^{(0)}}^\dag, & i {\bv^{(0)}}^t\\
V^t, & U^t\\
{\bu^{(0)}}^\dag, &  {\bu^{(0)}}^t\\
\end{pmatrix}.
}
Explicitly, the inverse transformation then becomes
\al{
\begin{split}
\label{EQ:QP_trafo_sigma_beta_insulator_homogeneous}
\tilde \sigma_\bk^{(i)}&=\sum_{\gamma}\left[ {u_{i}^{(\bk,\gamma)}} \, \beta_{\bk,\gamma} - {v_{i}^{(\bk,\gamma)}}^*  \,\beta_{-\bk,\gamma}^\dag \right]\\
 &+ \delta_{\bk,0} \left[  {v_{i,\bk}^{(0)}} \, \mathscr P +i {u_{i,\bk}^{(0)}} \, \mathcal Q \right]
\end{split}
\\
\begin{split}
\label{EQ:QP_trafo_sigma_beta_insulator_dag_homogeneous}
\tilde \sigma_\bk^{{(i)}^\dag}&=\sum_{\gamma}\left[ {u_{i}^{(\bk,\gamma)}}^* \, \beta_{\bk,\gamma}^\dag - {v_{i}^{(\bk,\gamma)}}  \,\beta_{-\bk,\gamma} \right]\\
&+ \delta_{\bk,0} \left[  {v_{i,\bk}^{(0)}}^* \, \mathscr P -i {u_{i,\bk}^{(0)}}^* \, \mathcal Q  \right].
\end{split}
}
In the MI the inverse transformation is analogous, up to the terms containing $\mP$ and $\mQ$ being absent.

\section{Expressing Operators in QP Operators}

To incorporate the effect of any external perturbation on the system or calculate expectation values, it is of central importance to express a given operator in terms of the new quasi-particle operators $\beta_{\bk,\gamma}^\dag$ and $\beta_{\bk,\gamma}$. Typically, an operator for our many-particle system is given in second quantization, in terms of momentum, quasi-momentum or Wannier creation and annihilation operators. It is a straightforward task to express such an operator in terms of local mean-field states in the form $\ket{i}_\bl {_\bl}\bra j$ and direct products thereof. Expressing a given operator in quasi-particle operators, the maximum number of $\beta_{\bk,\gamma}^\dag$ and $\beta_{\bk,\gamma}$ appearing in any term is twice as large as the maximum number of on-site operators appearing as a product in a given term. For instance, any on-site operator contains a maximum of two quasi-particle operator, any two site operator (such as the hopping term) a maximum of four, etc.. We now consider the transformation of operators of the form
\spl{
\label{EQ:FT_operator_A_type}
\mA_\bk=\frac{1}{\sqrt L} \sum_\bl e^{i \bk \cdot \bR_\bl} \tilde \mA_\bl,
}
where $\tilde A_\bl$ is an on-site operator. We define the matrix elements in the local Gutzwiller eigenbasis $A_{i,j}^{(\bl)}={_\bl}\bra i \tilde \mA_\bl \ket j _\bl$,
which become site independent for a homogeneous system and for the case that $\tilde \mA_\bl$ are the same local operator. Grouping by different orders in the fluctuations and using the fact that any local operator can be expressed in terms of at most two local fluctuation operators, we obtain
\spl{
\label{EQ:QP_operator_A}
\mA_\bk=& \sqrt L \, A_{0,0} \, \delta_{\bk,0} + \sum_{i>0}  \left( A_{i,0} \,  \sigma_{\bk}^{{(i)}^\dag}  +A_{0,i}  \,  \sigma_{-\bk}^{{(i)}}    \right) \\
&+\frac{1}{\sqrt L}\sum_{i,j>0}  \sum_{\bk'} ( A_{i,j}- A_{0,0}  \delta_{i,j}   ) \,  \sigma_{\lfloor \bk+\bk'\rfloor}^{{(i)}^\dag} \,  \sigma_{\bk'}^{{(j)}},
}
where the function $\lfloor \;  \rfloor:\bk \mapsto \lfloor \bk \rfloor = \bk+\bG$ maps a momentum vector into the first Brillouin zone (umklapp processes) by addition of an appropriate reciprocal lattice vector $\bG$ if $\bk$ lies outside the first Brillouin zone.

In the SF one obtains
\spl{
\label{EQ:Operator_rep_in_QP}
\mA_\bk
=& \sqrt L \, A_{0,0} \, \delta_{\bk,0} +\sum_{\gamma} \left[ D_{\bk,\gamma} \, \beta_{\bk,\gamma}^\dag +  \tilde D_{-\bk,\gamma}\, \beta_{-\bk,\gamma} \right]
\\&+D_{P}{\mathscr P} + D_{Q} \mathcal Q+\mathcal O(\beta^2),
}
where we defined the coefficients 
\al{
D_{\bk,\gamma}&=\sum_{i>0} \left[ A_{i,0} \, u_i^{(\bk,\gamma)^*}  -  A_{0,i} \, v_i^{(-\bk,\gamma)^*}\right]\\
\tilde D_{\bk,\gamma}&=\sum_{i>0} \left[   A_{0,i} \, u_i^{(\bk,\gamma)} - A_{i,0} \, v_i^{(-\bk,\gamma)}\right]\\
D_P&=\delta_{\bk,0} \sum_{i>0} \left[ A_{0,i} \, v_{i,-\bk}^{(0)}  +  A_{i,0} \, v_{i,\bk}^{(0)^*}   \right]\\
D_Q&=i \delta_{\bk,0} \sum_{i>0} \left[ A_{0,i} \, u_{i,-\bk}^{(0)}  -  A_{i,0} \, u_{i,\bk}^{(0)^*}   \right]
}
of the quasi-particle operators in the first order term. In the MI the transformation is identical, up to the last two terms (containing $\mathcal Q$ and $\mathcal P$) in Eq.~(\ref{EQ:Operator_rep_in_QP}) being absent.

For typical linear response calculations close to equilibrium where the density of quasi-particles is low,  the last term of second order in $\beta$ in Eq.~(\ref{EQ:Operator_rep_in_QP}) can be neglected. Also note that for Hermitian operators $\mathcal A_\bk$, the coefficients $D_P$ and $D_Q$ of the condensate mode position and momentum operators are purely real. The notation $\mathcal O(\beta^2)$ denotes terms of second order in any of the products of $\beta,\beta^\dag,\mathscr P$ and $\mathcal Q$.

Using this procedure, expectation values of arbitrary operators, such as the single-particle density matrix $\rho_{\bl,\bl'} = \ev{ b_\bl^\dag b_{\bl'} }$ and the quasi-momentum distribution can be explicitly evaluated.

\section{Completeness of $\sigma$-Coherent States}
The completeness relation can be proven by explicit integration of each fluctuation coherent state parameter in the complex plane in spherical coordinates $\phi_j=r_j e^{i \varphi_j}$. Hence, $\int d\Phi \, d\Phi^*=\prod_{j>0} \frac{ 1}{\pi}\int_0^\infty r_j \, dr_j  \int_0^{2\pi} d\varphi_j $ and defining $r_0=1$ and $\varphi_0=0$, one finds
\spl{
&\int d\Phi \, d\Phi^* \, e^{-\sum_{j>0} |\phi_j|^2 } \ket{\Phi}\bra{\Phi}\\
=&\left[ \prod_{j>0} \frac 1 \pi { \int_0^\infty r_j e^{-r_j^2} \, dr_j  \int_0^{2\pi} d\varphi_j } \right] \\
&\times \sum_{j',j''\geq 0} r_{j'} r_{j''} e^{i (\varphi_{j'}-\varphi_{j''})} \ket{j'}\bra{j''} \\
=& \sum_{j\geq 0}  \ket{j}\bra{j}=\mathbbm{1}, 
}
where we used $2\int_0^\infty r_j^{m} \, e^{-r_j^2} \, dr_j=1 $ for $m\in\{1,\,3\}$ and that the angular integral of any phase factor $e^{i (\varphi_{j'}-\varphi_{j''})}$ vanishes unless $j'=j''$.

\end{appendix}
\end{document}